\documentclass[12pt,a4paper]{iopart}
\usepackage{iopams}
\usepackage{graphicx}
\usepackage{color}
\usepackage{subfigure}

\begin{document}

\title[TRAPLESS BEC WITH TWO-AND THREE-BODY INTERACTIONS]{Stability of trapless Bose-Einstein condensates with two- and three-body interactions}

\author{S. Sabari, \footnote[1] {sabaripu@gmail.com} R.~Vasantha Jayakantha Raja, \footnote[2] {rvjraja@yahoo.com} K. Porsezian, \footnote[7]{ponz.phy@pondiuni.edu.in}}
\address{Department of Physics, Pondicherry University, Puducherry -- 605014, India}

\author{P. Muruganandam}

\address{School of Physics, Bharathidasan University, Tiruchirappalli -- 620024, India}

\begin{abstract}
We study the stabilization of a trapless Bose-Einstein condensate by analyzing the mean-field Gross-Pitaevskii equation with attractive two- and three-body interactions through both analytical and numerical methods. By using the variational method we show that there is an enhancement of the condensate stability due to the inclusion of three-body interaction in addition to the two-body interaction. We also study stability of the condensates in the presence of time varying three-body interaction. Finally we confirm the stabilization of a trapless condensates from numerical simulation.
\end{abstract}


\vspace{2pc}
\noindent{\it Keywords}: Bose-Einstein condensates, Three-body interaction, Variational approximation method, Crank-Nicholson method


\section{Introduction}
Bose-Einstein condensates (BECs) first realized experimentally in 1995 for rubidium~\cite{Anderson1995}, lithium~\cite{Bradley1995, Bradley1997}, and sodium~\cite{Davis1995}, provide unique opportunities for exploring quantum phenomena on a macroscopic scale. The properties of a condensate at absolute zero temperature are usually described by the time-dependent, nonlinear, mean-field Gross-Pitaevskii (GP) equation~\cite{Dalfovo1999}. The effect of the interatomic interaction leads to a nonlinear term in the GP equation. The s-wave scattering length, $a_s(t)$, plays an important role in the description of atom-atom interaction at ultralow temperatures $(T<1mK)$. The magnitude and sign of the s-wave scattering length, $a_s(t)$, can be tuned to any value, large or small, positive or negative by applying an external magnetic field. It is given by $a_s(t) = a \left[ 1 + \Delta/\left( B_0 - B(t) \right) \right]$, where $B(t)$ is the time-dependent externally applied magnetic field, $\Delta$ is the width of resonance and $B_0$ is the resonant value of the magnetic field. The presence of attractive interaction ($a_s(t)<0$) between the atoms has a profound effect on the stability of a BEC, since a large enough attractive interaction will cause the BEC to become unstable and collapse in some way. It is understood that at low temperature and density, where interatomic distances are much greater than the distance scale of atom-atom interactions, two-body interaction can be described by a single parameter (scattering length) where the effects of three-body interaction are negligible. At low enough temperatures the magnitude of the scattering length $a_s(t)$ is much less than the thermal de Broglie wavelength and the exact shape of the two-atom interaction is unimportant. On the other hand, if the atom density is considerably high the three-body interaction can start to play an important role~\cite{Abdullaev2001,Wamba2008,Ping2009}. For an attractive interatomic interaction the condensate is stable for upto a maximum critical number of atoms. When the number of atoms increases beyond this critical value, due to interatomic attraction, the radius of the BEC tends to zero and the maximum density of the condensate tends to infinity. With a supply of atoms from an external source the condensate can grow again and thus a series of collapses can take place, this has been observed experimentally in BEC of $^7${Li} with attractive interaction~\cite{Bradley1995, Bradley1997}. Theoretical analysis based on the GP equation also confirms the collapse. Thus for a system of atoms with attractive two-body interaction, the condensate has no stable solution above a certain critical number of atoms $N_{max}$~\cite{Edwards1995, Adhikari2000, Ruprecht1995}. However, as reported by Gammal et al~\cite{Gammal2000}, the addition of a repulsive potential derived from three-body interaction is consistent with a number of atoms larger than $N_{max}$. Even for a very small strength of the three-body interaction, the region of stability for the condensate can be extended considerably. By considering the possible effective interaction, it has been reported that a sufficiently dilute and cold Bose gas exhibits similar three-body dynamics for both signs of the s-wave scattering length~\cite{Ercy1996}. It was also suggested that, for a large number of bosons the three-body repulsion can overcome the two-body attraction, and a stable condensate will appear in the trap~\cite{Josser1997}.  It is worth to mention that Ping et al. have studied the two- and three-body interactions through analytical studies in a trapped BEC using the so called Gross-Pitaevskii-Ginzburg equation~\cite{Ping2009}.

Weakly interacting BECs atoms have stimulated intensive interest in the field of atomic matter waves and nonlinear excitations such as dark \cite{Anderson2001, Denschlag2000} and bright solitons \cite{Strecker20002, Khaykovich2002,Wuster2009}. A numerical study of the time-dependent GP equation is of interest, as this can provide solutions to many stationary and time-evolution problems. The time-independent GP equation yields only the solution of stationary problems. As our principal interest is in time evolution problems, we shall only consider the time-dependent GP equation in this paper.  As the problem of the stabilization of a soliton in a trapless condensate is of utmost interest in several areas, for example, nonlinear physics~\cite{Kivshar2003}, optics~\cite{Kivshar2003, Towers2002} and BECs, in the present study we reexamine the problem of stabilization and point out that a temporal modification of the scattering length can lead to a stabilization of the trapless soliton in three dimensions. In this paper, in addition to analytical studies, we also perform numerical verification for the stability of trapless BEC in the presence of three-body interaction. In particular, by analyzing the GP equation using variational method and direct numerical integration, we address stabilization properties in most of the possible cases where the two- and three-body interactions can be realized. Our present analysis strongly suggests that the inclusion of three-body interaction of suitable form can stabilize the trapless BEC. We also illustrate from numerical simulations that the untrapped attractive condensate can maintain a reasonably constant spatial profile over a sufficient interval of time through temporal modulation.

The organization of the present paper is as follows. In Section~\ref{sec2}, we present a brief overview of the mean-field model. Then, we discuss the variational study of the problem and point out the possible stabilization of a trapless BEC with three-body interaction in Section~\ref{sec3}. In Section~\ref{sec4}, we report the numerical results of the time-dependent GP equation with two- and three-body interactions through split-step Crank-Nicholson (SSCN) method and we investigate the stability of a trapless BEC for two different cases. Finally, we give the concluding remarks in Section~\ref{sec5}.

\section{Nonlinear mean-field model}
\label{sec2}

At ultra low temperatures the time dependent wave function of the condensates $\Psi(\tilde{\textbf{r}},\tau)$ at position $\tilde{\textbf{r}}$ and time $\tau$ in the presence of three-body interaction can be described by the following mean-field nonlinear GP equation
\cite{Ping2009,Gammal2000,Fetter1971}
\begin{eqnarray}
\fl \left[ -i\hbar\frac{\partial}{\partial \tau}-\frac{\hbar^2\nabla^2}{2m}+ V(\tilde{\textbf{r}})+ g(\tau) N|\Psi(\tilde{\textbf{r}},\tau)|^2+ k(\tau) N^2 |\Psi(\tilde{\textbf{r}}, \tau)|^4 \right]\Psi(\tilde{\textbf{r}},\tau) = 0, \label{Jac1}
\end{eqnarray}
where $N$ is the number of atoms in the condensate, $V(\tilde{\textbf{r}})=m\omega^2 \tilde{r}^2/2$, is the spherically symmetric trap geometry, $g(\tau)=4\pi \hbar^2 a_s(\tau)/m $ and $k(\tau)$ are the strengths of time dependent two-body and three-body interatomic interactions, respectively. $\hbar$ is Planck's constant and $m$ is mass of the single bosonic atom, $a_s(t)$ is time dependent s-wave scattering length which can be tuned to any desired value by using Feshbach resonance technique.  The normalization condition is $\int |\Psi(\tilde{\textbf{r}}, \tau)|^2 d\tilde{\textbf{r}}  = 1$. Usually the strength of the three-body interaction is very small when compared with strength of the two-body interaction as pointed out by Gammal~\cite{Gammal2000}. 
Accordingly we have considered $k(\tau)\approx 10$ percent of $g(\tau)$ for our present study. It may be noted that, since $k(\tau)$ is function of $g(\tau)$, the three body interactions can also be controlled by the tuning of s-wave scattering length~\cite{Gammal2000}. In the present work, we essentially look for stabilization of trapless BEC. When one expects solitons in BEC using GP equation, the system should be conservative. It means that the GP equation should not have any dissipative term like gain/loss etc. If we include the effect of gain/loss of atoms then the corresponding GP equation will be a non-conservative system and hence there is no soliton in the conventional sense. However, one can still look for non-autonomous solitons by suitably tailoring the gain/loss of atoms. For example, such non-autonomous solitons have been studied by Rajendran et al.~\cite{Rajendran2009,Rajendran2010}, Serkin et al.~\cite{Serkin2007,Serkin2010}. The nature of such solitons in the case with both two- and three-body interactions and gain/loss of atoms have been considered to some extent by Roy et al.~\cite{Roy2010}. However, in the present study, we mainly focus on the stabilization of trapless BEC for conservative system.

It is more convenient to use the GP equation (1) into a dimensionless form. For this purpose we make the transformation of variables as $r = \sqrt{2}\tilde{r}/l$, $t = \tau\omega$, $l = \sqrt{\hbar/(m\omega)}$ and  $\phi(r,t)=\Psi(\tilde{\textbf{r}},\tau)(l^3/2\sqrt{2})^{1/2}$. Then, the radial part of the GP equation (\ref{Jac1}) becomes \cite{Adhikari2004,Muruganandam2009},
\begin{eqnarray}
\fl \left[{-i\frac{\partial}{\partial t}}- \left(\frac{\partial^2}{\partial r^2}+\frac{D-1}{r}\frac{\partial}{\partial r}\right)+ \frac{r^2}{4}d(t)+{ g(t)\left\vert\phi(r,t)\right\vert^2}+\chi(t) \left\vert\phi(r,t)\right\vert^4 \right]\phi(r,t) = 0, \label{Jac2}
\end{eqnarray}
here $D$ represents a spatial dimension,
The parameter $d(t)$ represents the strength of the external trap which is to be reduced from $1$ to $0$  when the trap is switched off. The normalization condition in this case is $4\pi\int_{0}^{\infty} |\phi(r,t)|^2 dr= 1$.

\section{Variational approximation}
\label{sec3}

\begin{figure}[!ht]
\begin{center}
\includegraphics[width=.45\textwidth]{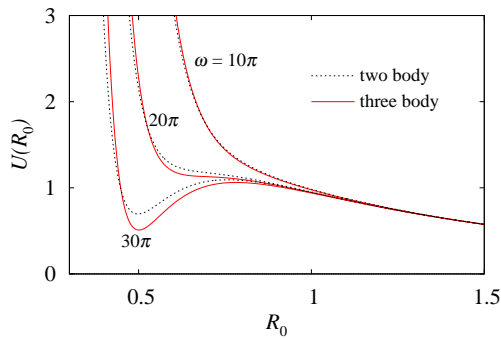}
\end{center}
\caption{Plot of the effective potential $U(R_0)$ of equation~(\ref{Jac13}) as a function of $R_0$ with (solid line) and without (dotted line) three-body interaction for different $\omega$ values and for $g_0 = -25$.}
\label{f1}
\end{figure}

In the following, we use the variational approach with the trial wave function (Gaussian ansatz) for the solution of equation (\ref{Jac2}) where the external potential is absent~\cite{Adhikari2004,Adhikari2001}:
\begin{eqnarray}
\phi(r,t)=N(t)\exp{\left[-\frac{r^2}{2R(t)^2}+\frac{i}{2} \beta(t)r^2 +i \alpha(t)\right]}, \label{Jac3}
\end{eqnarray}
where, $N(t)= [ \pi^{\frac{3}{4}}  R(t)^{\frac{3}{2}}]^{-1}$ for D:3 and $N(t)=[ \sqrt{\pi}R(t)]^{-1}$ for D:2, $R(t)$, $\beta(t)$  and  $\alpha(t)$ are the normalization, width, chirp and phase of the system, respectively. The Lagrangian density for equation (\ref{Jac2}) is given by
\begin{eqnarray}
\fl \mathcal{L} = \frac{i}{2}\left(\frac{\partial\phi}{\partial t}\phi^*-\frac{\partial\phi^*}{\partial t}\phi\right)r^{D-1}-\left|\frac{\partial\phi}{\partial r}\right|^2 r^{D-1}-\frac{r^{D-1}}{2}g(t)|\phi|^4-\frac{r^{D-1}}{3}\chi(t)|\phi|^6. \label{Jac4}
\end{eqnarray}
The trail wave function equation (\ref{Jac3}) is substituted in the Lagrangian density and the effective Lagrangian is calculated by integrating the Lagrangian density as $L_{eff} = \int \mathcal{L}\, dr$. The Euler-Lagrangian equations for $R(t)$ and $\beta(t)$ are then obtained from the effective Lagrangian in a standard fashion as,
\begin{eqnarray}
\dot{R}(t) & = & 2 R(t)\beta(t),\label{Jac6} \\
\dot{\beta}(t) & = & \frac {2} {R(t)^4} - 2 \beta(t)^2 + \frac{g(t)}{2\sqrt{2 \pi^3} R(t)^5}+\frac{4\chi(t)}{9\sqrt{3} \pi^3 R(t)^8}.\label{Jac7}
\end{eqnarray}
By combining the equations (\ref{Jac6}) and (\ref{Jac7}), we get the following second-order differential equation for the evolution of the width,
\begin{eqnarray}
\ddot{R}(t) = \frac {4}{R(t)^3} + \frac{g_0 + g_1 \sin{(\omega t)}}{\sqrt{2 \pi^3} R(t)^4}+\frac{8( \chi_0 +\chi_1 \sin{(\omega t)})}{9\sqrt{3} \pi^3 R(t)^7}, \label{Jac8}
\end{eqnarray}
with $ g(t) = g_0 + g_1 \sin{(\omega t)}$ and $\chi(t) =  \chi_0 +\chi_1 \sin{(\omega t)}$, where $g_0 $, $\chi_0 $ are  constant part of the scattering length  of two-body, three-body interaction respectively and $g_1$, $\chi_1$ are the amplitude of oscillating part of the scattering length. Now $R(t)$ can be separated into a slowly varying part $R_0(t)$ and a rapidly varying part $\rho(t)$ by $R(t)=R_0(t)+\rho(t)$. When $ \omega\ \gg 1 $, $\rho(t)$ becomes of the order of $\omega^{-2}$. Keeping the terms of the order of up to $\omega^{-2}$ in $\rho(t)$, one may obtain the following equations of motion for $R_0(t)$ and $\rho(t)$~\cite{Landau1960},
\begin{equation}
\ddot{\rho}(t) = \frac{ g_1 \sin{(\omega t)}}{\sqrt{2 \pi^3} R_0(t)^4}+\frac{8\chi_1 \sin{(\omega t)}}{9\sqrt{3} \pi^3 R_0(t)^7},\label{Jac9}
\end{equation}
\begin{equation}
\fl \ddot{R_0}(t) = \frac {4}{R_0^3(t)} + \frac{g_0}{\sqrt{2 \pi^3} R_0^4(t)}+\frac{8\chi_0}{9\sqrt{3}\pi^3 R_0^7(t)}- \frac{4g_1  \overline{\rho(t) \sin(\omega t)}}{\sqrt{2\pi^3} R_0^5(t)} -\frac{56 \chi_1  \overline{\rho(t) \sin(\omega t)}}{9\sqrt{3}\pi^3 R_0^8(t)},\label{Jac10}
\end{equation}
where the overline indicates the time average of the rapid oscillation. From equation (\ref{Jac9}) we can get $\rho(t)$ and substituting it into equation (\ref{Jac10}), we obtain the following equation of motion for the slowly varying part,
\begin{eqnarray}
\fl \ddot{R_0} = \frac {4}{R_0^3} + \frac{g_0}{\sqrt{2 \pi^3} R_0^4}+\frac{8\chi_0}{9\sqrt{3}\pi^3 R_0^7}+\frac{g^2_1}{\pi^3 \omega^2 R_0^9}+\frac{22\sqrt{2} g_1\chi_1}{9\pi^4\sqrt{3\pi}\omega^2 R_0^{12}}+\frac{224 \chi^2_1}{243\pi^6\omega^2 R_0^{15}},\label{Jac12}
\end{eqnarray}
and the effective potential $U(R_0)$ corresponding to the above equation of motion can be written as,
\begin{eqnarray}
\fl U(R_0) =\frac {2}{R_0^2} + \frac{g_0}{3\sqrt{2 \pi^3} R_0^3}+\frac{4\chi_0}{3^{7/2}\pi^3 R_0^6}+ \frac{g^2_1}{8 \omega^2\pi^3 R_0^8 }+\frac{2\sqrt{2} g_1\chi_1}{9\pi^4\sqrt{3\pi}\omega^2 R_0^{11}}+\frac{16 \chi^2_1}{3^5\omega^2\pi^6 R_0^{14}}, \label{Jac13}
\end{eqnarray}
 If one considers the two-body interaction alone, that is, $\chi_0 = 0$ and $\chi_1  = 0$, the effective potential can be reduced as,
\begin{eqnarray}
U(R_0) =\frac {2}{R_0^2} + \frac{g_0}{3\sqrt{2 \pi^3} R_0^3}+ \frac{g^2_1}{8 \omega^2\pi^3 R_0^8 }.\label{Jac14}
\end{eqnarray}
 which is exactly the same as discussed in ref \cite{Adhikari2004}. Now we analyze the nature of the effective potential in the presence and in the absence of three-body interaction. Figure~\ref{f1} depicts the potential energy
curves as a function of $R_0$ in the absence and in the presence of three-body interaction for different frequencies of the periodic force. One may infer from Figure 1 that for the potential energy curve does not show any minimum $\omega = 10\pi$ in order to have a stable condensates. On increasing the frequency to $\omega = 30\pi$, a minimum dip appears in the potential for both the cases. It is also evident that the inclusion of three-body interaction deepens the minimum as represented by the solid line in Figure~\ref{f1}. We consider the case with $\omega=30\pi$ for further analysis. Next we look into the stability of the condensate upon varying the nonlinearity $g_0$ with two-body interaction alone. Thus the inclusion of three-body interaction seems to increase the stability of the condensates.
\begin{table}[!ht]
\caption{Results of variational approximation}
\label{table1}
\begin{center}           
\begin{tabular}{|c|l|r|}    
\hline \hline                   
	\multicolumn{1}{|c|}{Case}
        & \multicolumn{1}{|c|}{Type of Interactions}
	& \multicolumn{1}{|c|}{$g_0$ (critical)} \\
\hline \hline                                
a & Two-body interaction (constant and oscillation)  & -21.4527 \\      
b & Case a and three-body interaction (constant and oscillation) & -20.2774  \\
c & Case a and three-body interaction (constant only) & -18.9403 \\
d & Case a and three-body interaction (oscillation only) & -22.3423  \\
\hline \hline         
\end{tabular}
\end{center}
\end{table}

The stability of trapless BEC with two-body interaction for constant (slowly varying) and oscillatory (rapidly varying) part has been already explored~\cite{Adhikari2004,Abdullaev2003,Saito2003}. However, to the best of our knowledge, the effect on the inclusion of three-body interaction has not been studied in trapless BEC. Hence, in the present study, we are interested to analyze the effect of three-body interaction on the stability of trapless BEC. To analyze the effect of three-body interaction, we consider four types of different possible combinations of two- and three-body interactions as mentioned in Table~\ref{table1}, namely, (a) two-body interaction alone (both constant and oscillatory part),
\begin{figure}[!ht]
\begin{center}
\includegraphics[width=.45\textwidth]{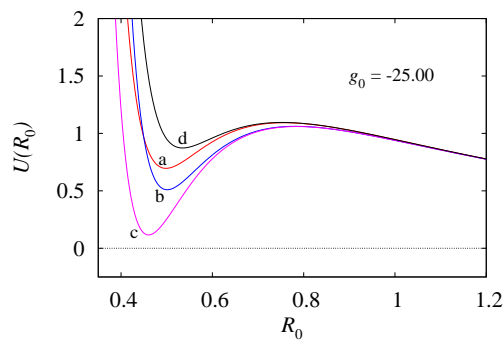}
\end{center}
\caption{The effective potential $U(R_0)$ versus $R_0$ and $g_0 = -25$. Curve (a) indicates the case in the two-body interaction alone. Curve (b) indicates the presence of two- and three-body interactions. Curve (c) indicates the two-body interaction with the presence of the constant part of the three-body interaction only. Curve (d) indicates the two-body interaction with the absence of the constant part of the three-body interaction.}
\label{f4}
\end{figure}(b) two-body interaction (case a)  with constant and oscillatory form of three-body interactions,(c) two-body interaction (case a) with constant three-body interaction and (d) two-body interaction (case a) with oscillatory three-body interaction. In Figure~\ref{f4}, we plot the potential energy curves as a function of distance for different types of interaction for a fixed value of $g_0 = -25$ and $\omega = 30\pi$. Curve (a) in Figure~\ref{f4} is drawn by considering the two-body interaction alone, curve (b) represents the variation of potential energy in the case of two-body interaction with both constant and oscillatory form of three-body interactions, curve (c) illustrates the potential energy in the case of two-body interaction with constant three-body interaction and in curve (d) we show the  potential energy for the case of two-body interaction with oscillatory three-body interaction. It is evident from Figure~\ref{f4} that the inclusion of constant three-body interaction [curve (c)] has the maximum depth in the potential energy. The critical values of $g_0$ below which the condensate is stable for the above four cases is given in Table~\ref{table1}. We have also studied stability of the trapless BEC for the cases (a) and (c) by numerically solving the variational  equations~(\ref{Jac10}). In Figure~\ref{f2}(a) we show the potential energy for different $g_0$ values by considering two-body interaction alone.
\begin{figure}[!ht]
\begin{center}
\includegraphics[width=0.85\textwidth]{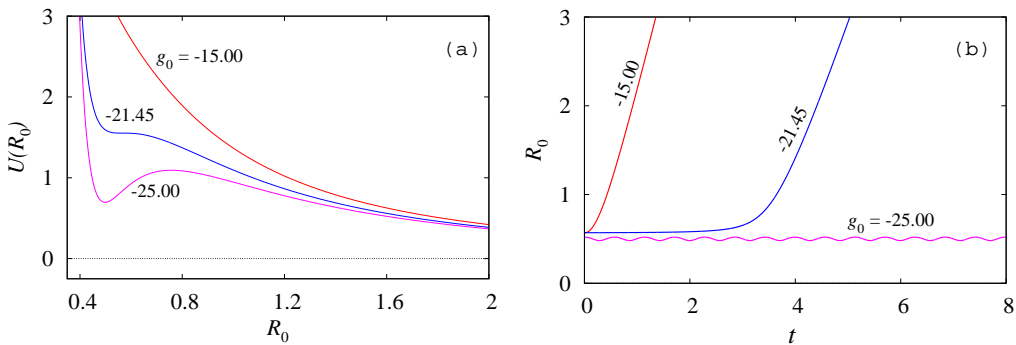}
\end{center}
\caption{(a) Plot of the effective potential $U(R_0)$ versus $R_0$ and (b) the equilibrium width $R_0$  as a function of time for different  $g_0$ in equation~(\ref{Jac14}).}
\label{f2}
\end{figure}
Since the stability of case (d) is very low (minimum depth in the potential curve d in Figure~\ref{f2}) than two-body interaction and the stability of case (b) is lower than case (c), we have considered the role of three-body interaction in the presence of constant part with two-body interaction only. In order to compare the influence of three-body interaction, we have also considered a case with two-body interaction alone [case (a)]. The variation of effective potential and effective width of two-body and three-body interactions of trapless BEC are shown in Figure~\ref{f2} and Figure~\ref{f3}. Figure~\ref{f2}(a) and Figure~\ref{f2}(b) depict the role of effective potential for various $g_0$ values and the dynamics of size of the condensates for corresponding values respectively. We have observed in Figure~\ref{f2}(a) that there is no potential depth for $g_0 = -15.00$. Hence the system becomes weakly attractive and the condensates expand to infinity. Also, when negative $g_{0}$ value increases, we have observed the potential depth at $g_0 = -21.45$ which is called critical depth. It is clearly shown the same in Figure~\ref{f2}(b) that the size of condensates  stable up to three time units and  in the final stage it eventually collapses. If $g_0$ increase to -25.00, the depth of the minimum in the effective potential increased. It means, the system becomes highly attractive  and the size of condensates are stable for long  time units and  in the final stage it may collapse. The role of three-body interaction of trapless BEC is illustrated in Figure~\ref{f3}.  From Figures~\ref{f2} and \ref{f3} we noted that one can obtain critical value at the minimum $g_{0}$ value in the presence of three-body interaction when compared to two-body interaction.

\begin{figure}[!ht]
\begin{center}
\includegraphics[width=0.85\textwidth]{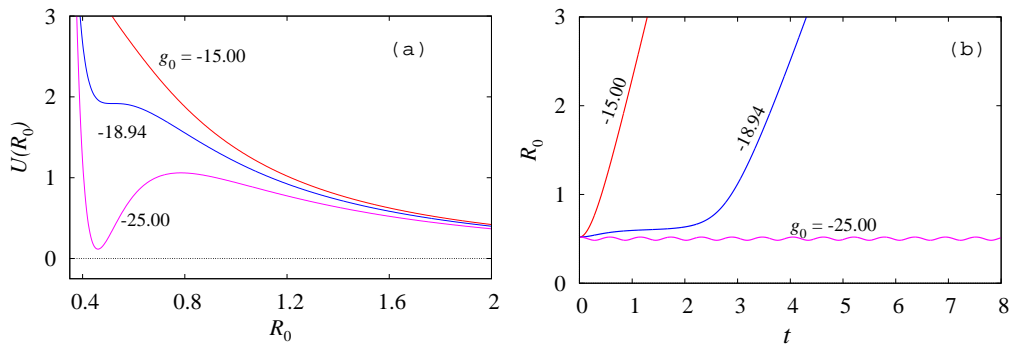}
\end{center}
\caption{Figure (a) showing the effective potential $U(R_0)$ versus $R_0$ and (b) showing width ($R_0$) as function of time ($t$) by from the numerical solution of equation~(\ref{Jac13}) for different values of $g_0$. Here  $g_1  = -4\,g_0$, $\chi_0 = 0.1\,g_0$ and $\chi_1 = 0$.}
\label{f3}
\end{figure}
\section{Numerical Results}
\label{sec4}
Next we study the stability properties of trapless BEC by solving the time-dependent GP equation~(\ref{Jac2}) numerically through SSCN method \cite{Muruganandam2009,Adhikari2002}. For this purpose, we transform the wave function $\psi(r,t)/r = \phi(r,t)$ and choose the boundary condition of the wave function as $r\to0$ to $\infty$. Hence the cubic and quintic nonlinear term can eventually be neglected in the GP equation for large $r$ and equation~(\ref{Jac2}) becomes,
\begin{eqnarray}
\fl \left[{-i\frac{\partial}{\partial t}}- \frac{\partial^2}{\partial r^2}+\frac{r^2}{4} d(t)+{ g(t)\left\vert\frac{\psi(r,t)}{r}\right\vert^2}+\chi(t) \left\vert\frac{\psi(r,t)}{r}\right\vert^4 \right]\psi(r,t) = 0, \label{Jac15}
\end{eqnarray}
where, $g(t)=g_f[a_1-b_1\sin(\omega t)]$ and $\chi(t)=\chi_f[a_2-b_2\sin(\omega t)]$ are the strength of the two- and three-body interactions respectively. Here, the set of parameters $g_f$, $a_1$, $b_1$ and $\chi_f$, $a_2$, $b_2$ correspond to final, constant and co-efficient of oscillatory part of two- and three-body interactions, respectively.
To solve the GP equation for large nonlinearity $\vert g(t) \vert$ and $\vert \chi(t) \vert$, one may start with the Thomas-Fermi approximation for
\begin{figure}[!ht]
\begin{center}
\includegraphics[width=0.85\textwidth]{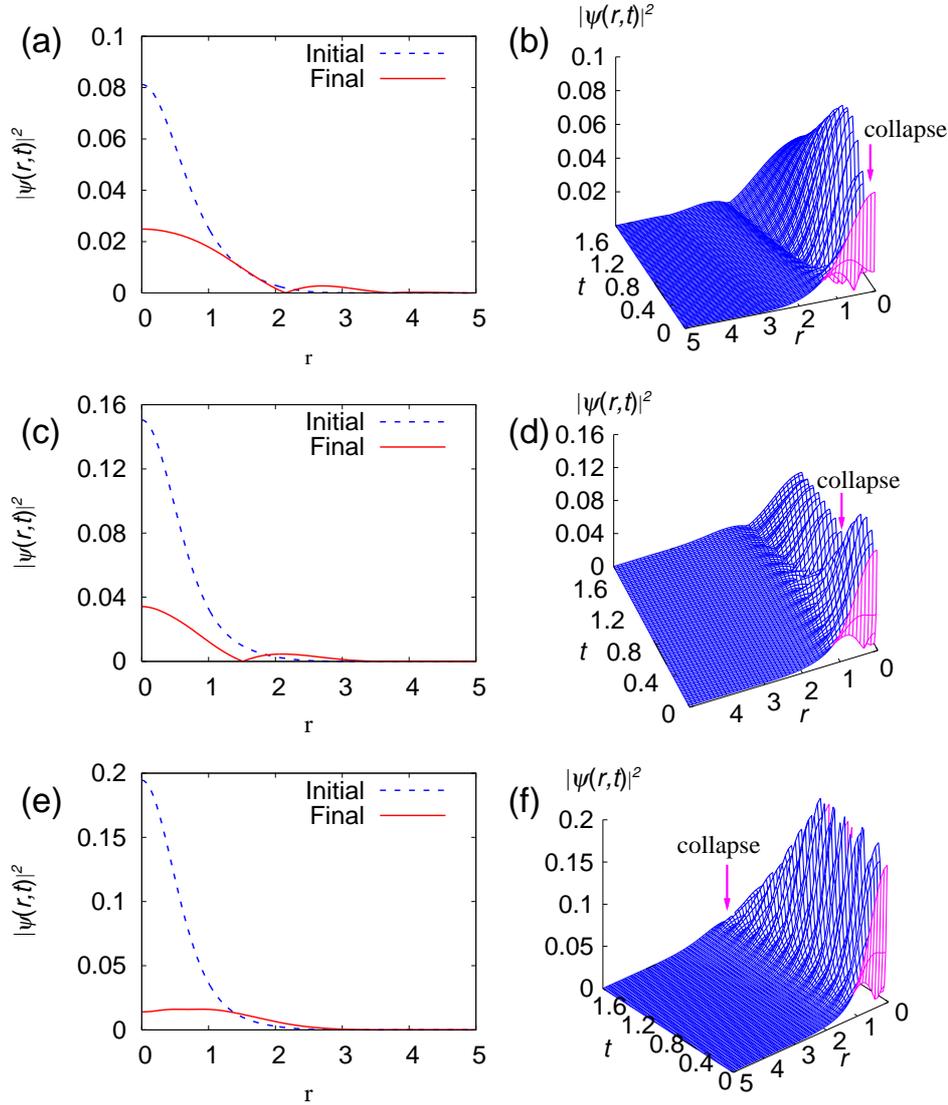}
\end{center}
\caption{Stabilization of trapless BEC in the presence of two-body interaction alone (case a) $\chi(t)=0$ and $a_1=1$, $b_1=4$ in Equation~(\ref{Jac15}), $g_f=-15.00$ [(a) and (b)], $g_f=-21.45$ [(c) and (d)] and $g_f=-25.00$ [(e) and (f)]}
\label{f5}
\end{figure}
\begin{figure}[!ht]
\begin{center}
\includegraphics[width=0.85\textwidth]{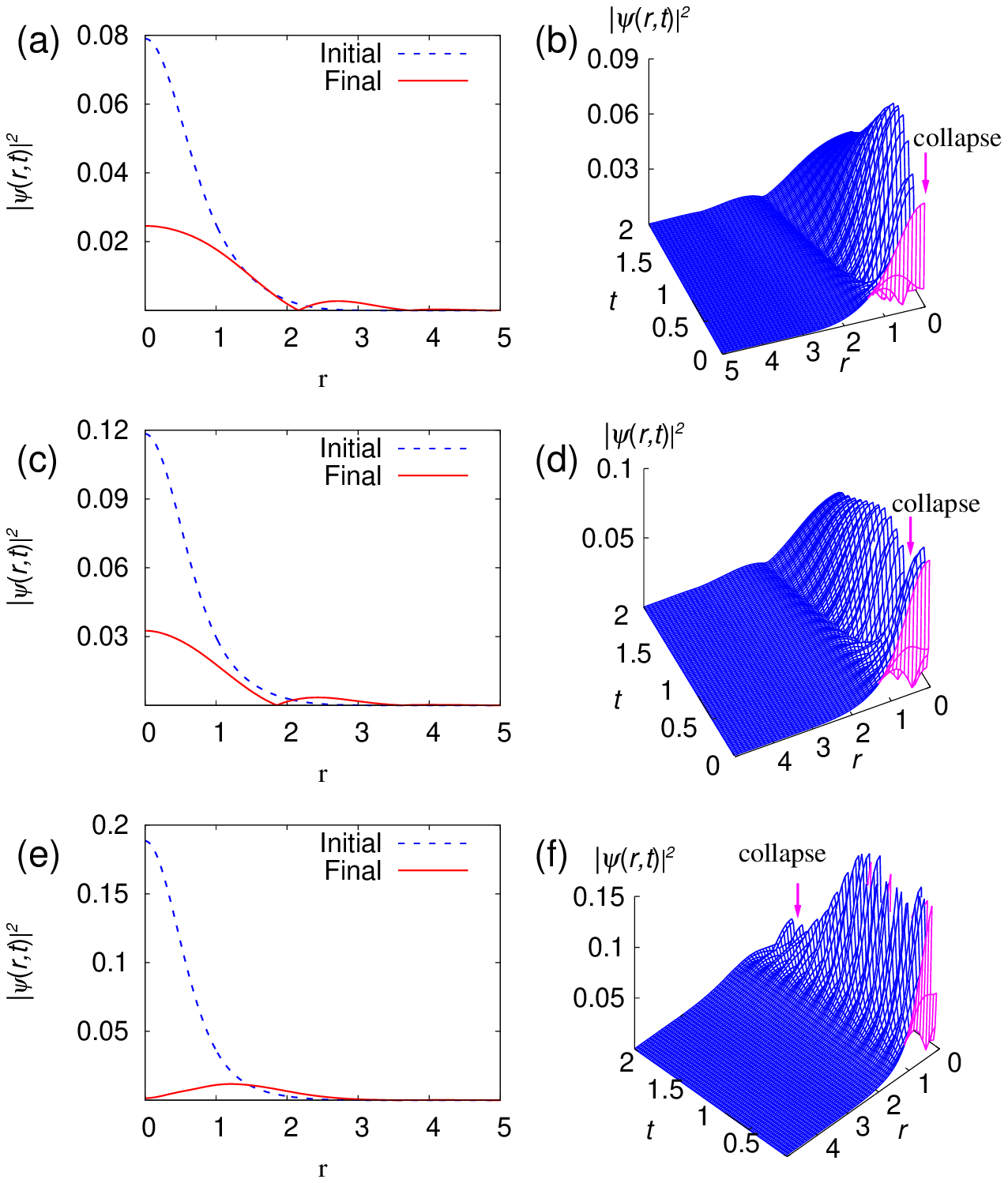}
\end{center}
\caption{Stabilization of trapless BEC in the presence of two- and three-body interaction (case c) $\chi_f=0.1g_f$ and $a_1=a_2=1$, $b_1=4$, $b_2=0$ in Equation~(\ref{Jac15}), $g_f=-15.00$  [(a) and (b)], $g_f=-18.94$ [(c) and (d)] and $g_f=-25.00$ [(e) and (f)]}
\label{f6}
\end{figure}
the wave function obtained by setting all the derivatives in the GP equation to zero, which is a good approximation for large nonlinearity \cite{Dalfovo1999,Muruganandam2009,Thogersen2009,Fetter2009}. Alternatively, the harmonic oscillator solution is also a good starting point for small values of nonlinearity as in this paper. The typical discretized space and time steps for solving SSCN method is 0.01 and 0.0001. Then in the course of time iteration, the coefficient of the nonlinear term is increased from 0 at each time step. Simultaneously, the initial stage of harmonic trap is also switched off slowly by changing $d(t)$ from $1$ to $0$ until the final value of nonlinearity attained at a certain time called time $t_0$. Because, one needs to reduce the harmonic trap frequency while increasing the nonlinearity for obtaining the stability. Otherwise, the trapping frequency will reduce the size of the condensate, may collapse due to attraction. During this process the harmonic trap is removed, and after the $g_f$, $\chi_f$ are attained at time $t_0$, the periodically oscillating nonlinearity $g(t)=g_f[a_1-b_1\sin(\omega t)]$ and $\chi(t)=\chi_f[a_2-b_2\sin(\omega t)]$ are applied for $t>t_0$~\cite{Adhikari2004,Saito2003}.

To investigate the stability of condensate in the presence of three-body  with two-body interactions using numerical simulation, we consider the crucial cases (a) and (c) only from the Table~\ref{table1}. Figure(5a), (5c) and (5e) illustrate the dynamics of two-body interaction (case a) only for different values of $g_f$ by setting $-15.00, -21.45, -25.00$ in equation~(\ref{Jac15}). The space-time plot of the density $|\psi(r,t)|^2$ is shown in Figure(5b), (5d) and (5f).  The dominant physical parameters using for numerical simulations are $a_1=1$, $b_1=4$ and $\chi(t)=0$. It is noteworthy from Figure~\ref{f5}, although the peak density oscillates with respect to time due to oscillation nonlinearity, the density remains stable without breaking. Hence, the splitting of density profile is represented as collapse of the condensation. As seen from Figure~\ref{f2}, it is observed from Figure~\ref{f5} that one can increase the stability of condensates by increasing the negative value of $g_f$. The variation of density profile for three-body interaction with two-body (case c) for different $g_f$ is predicted in Figure~\ref{f6}. In this case, we have used the physical parameters as $a_1=a_2=1$, $b_1=4$, $b_2=0$, $\chi_f=0.1g_f$ and $g_f=-15.00, -18.94, -25.00$ for solving equation~(\ref{Jac15}) numerically. Since the value of final nonlinearity of three-body interaction value is very low when compared with two-body, we have considered the value of $\chi_f$ as 10 percentage of two-body nonlinearity value. It is clearly shown from Figure~\ref{f6} that the analytical solution of Figure~\ref{f3} is verified through numerical simulation. Hence we concluded that the stability of the condensation can be increased by considering the three-body interaction.

\section{Conclusion}
\label{sec5}
In conclusion, we have theoretically investigated the stabilization of trapless BEC using GP equations with two-and three body interactions. Before investigating the importance of three-body interaction in terms of stabilization, we have performed VA analysis and derived the equation of motion to investigate the stability of trapless BEC. Based on the analytical results, we have studied that the addition of three-body interation with two-body interaction, increases the stability of the system. We also analyzed different cases of interactions with presence/absence of constant/oscillatory part of the three-body interactions with two-body interaction. We also verified our analytical results with numerical simulation using SSCN method.  The numerical results exactly match with the results obtained by VA method. From our analytical and numerical results, it is clear that one can increase the stability of the trapless BEC by the inclusion of three-body interaction.

\ack
K.P. thanks to the DST-DFG, DST, DAE-BRNS, and UGC, Government of India, for the financial support through major projects. The work of PM is supported by DST, Government of India in the form of research project.

\section*{References}

\providecommand{\newblock}{}

\end{document}